\def\papertitle{Quality Audio Prototyping: a prototype system for unified sound retrieval and procedural generation.}
\def\paperauthorA{Nelly Garcia}
\def\paperauthorB{Aditya Bhattacharjee}
\def\paperauthorC{Gabryel Mason-Williams}
\def\paperauthorD{Israel Mason-Williams}
\def\paperauthorE{Emmanouil Benetos}
\def\paperauthorF{Joshua Reiss}

\documentclass[twoside,a4paper]{article}
\usepackage{etoolbox}
\usepackage{booktabs}

\usepackage{amsmath,amssymb,amsfonts,amsthm}
\usepackage{siunitx}
\usepackage{euscript}
\usepackage[T1]{fontenc}
\usepackage[utf8]{inputenc}
\usepackage{ifpdf}
\usepackage[english]{babel}
\usepackage{caption}
\usepackage{subfig} 
\usepackage{color}
\usepackage{booktabs}
\usepackage{multirow}
\usepackage{lipsum}
\usepackage[print]{dafx26v2}

\input glyphtounicode
\ninept

\newcounter{numauth}
\newcounter{listcnt}
\newcommand\authcnt[1]{\ifdefined#1 \stepcounter{numauth} \fi}

\newcommand\addauth[1]{
\ifdefined#1 
\stepcounter{listcnt}
\ifnum \value{listcnt}<\value{numauth}
\appto\authorslist{, #1}
\else
\appto\authorslist{~and~#1}
\fi
\fi}
\authcnt{\paperauthorB}
\authcnt{\paperauthorC}
\authcnt{\paperauthorD}
\authcnt{\paperauthorE}
\authcnt{\paperauthorF}
\authcnt{\paperauthorG}
\authcnt{\paperauthorH}
\authcnt{\paperauthorI}
\authcnt{\paperauthorJ}
\def\authorslist{\paperauthorA}
\addauth{\paperauthorB}
\addauth{\paperauthorC}
\addauth{\paperauthorD}
\addauth{\paperauthorE}
\addauth{\paperauthorF}
\addauth{\paperauthorG}
\addauth{\paperauthorH}
\addauth{\paperauthorI}
\addauth{\paperauthorJ}

\usepackage{times}

\newif\ifpdf
\ifx\pdfoutput\relax
\else
   \ifcase\pdfoutput
      \pdffalse
   \else
      \pdftrue
   \fi
\fi

\ifpdf 
  \usepackage[pdftex,
    pdftitle={\papertitle},
    pdfauthor={\authorslist},
    pdfsubject={Proceedings of the 29th International Conference on Digital Audio Effects (DAFx26)},
    colorlinks=false, 
    bookmarksnumbered, 
    pdfstartview=XYZ 
  ]{hyperref}
  \usepackage[pdftex]{graphicx}
\else 
  \usepackage[dvips]{epsfig,graphicx}
  \usepackage[dvips,
    pdftitle={\papertitle},
    pdfauthor={\authorslist},
    pdfsubject={Proceedings of the 29th International Conference on Digital Audio Effects (DAFx26)},
    colorlinks=false, 
    bookmarksnumbered, 
    pdfstartview=XYZ 
  ]{hyperref}
\fi
\usepackage[hypcap=true]{caption}
\title{\papertitle}

\affiliation
{\paperauthorA, \paperauthorB, \paperauthorC, \paperauthorD, \paperauthorE, \paperauthorF}
{\href{https://dafx26.mit.edu}{Queen Mary University of London, King's College of London} \\ London, UK\\
{\tt \href{mailto:n.v.a.garcia-sihuay@qmul.ac.uk}{n.v.a.garcia-sihuay@qmul.ac.uk}}
}

\begin{document}
\ifpdf 
  \DeclareGraphicsExtensions{.png,.jpg,.pdf}
\else  
  \DeclareGraphicsExtensions{.eps}
\fi


\maketitle

\begin{abstract}
Sound design workflows frequently oscillate between time-consuming library searches and the complexity of procedural synthesis, with practitioners typically relying on disconnected tools to address each challenge separately. This paper introduces Quality Audio Prototyping (QuAP), a working prototype that unifies content-based audio retrieval and procedural sound generation within a single interface, reducing the procedural distance between a narrative concept and its sonic realisation. QuAP integrates a similarity-based retrieval engine with real-time procedural audio models, complemented by a rule-based assistant that provides perceptually informed parameter guidance, offering definitions and recommendations derived from empirical optimisation rather than requiring prior synthesis knowledge. Preliminary evaluation confirms the viability of this approach: subjective assessment demonstrated statistically significant quality improvements in five of six embedded synthesis models, and an encoder ablation study established the preferred retrieval architecture on a sound effect dataset. A user evaluation with 16 practitioners confirmed the tool's workflow utility, with all participants agreeing that the parameter assistant preserved creative agency while lowering the barrier to procedural interaction.
\end{abstract}

\section{Introduction}
\label{sec:intro}

Sound design encompasses the creation and manipulation of environmental sounds, layered textures, and sound elements that compose the auditory dimension of audiovisual productions, distinct from dialogue and music~\cite{sonnenschein2001sound}. To create a soundscape, practitioners rely on extensive sound libraries, Foley~\cite{ament2009foley} recordings, and production stems, navigating  large-scale collections under time and creative pressure.

The sound design workflow combines technical and aesthetic skills in equal measure. On the technical side, tasks such as sound retrieval, source separation, and parameter manipulation represent recurring bottlenecks where signal processing and machine learning approaches offer potential for optimisation~\cite{Kamath}. However, existing tools typically address these bottlenecks in isolation: retrieval systems, synthesis engines, and organisational tools operate as disconnected applications, requiring practitioners to context-switch between environments and interrupting creative flow during the exploratory phase of a project.

\begin{table}
\centering
\caption{Comparison between traditional sound design 
workflows and the proposed QuAP system.}
\label{tab:workflow_comparison}
\resizebox{\linewidth}{!}{
\begin{tabular}{@{}lll@{}}
\toprule
\textbf{Feature} & \textbf{Traditional} & \textbf{QuAP (Proposed)} \\ \midrule
Search Method & Text / Metadata & Hybrid (Text + Embeddings) \\
Retrieval Speed & Manual browsing & FAISS-accelerated vector search \\
Sound Creation & External plugins & Integrated procedural models \\
Workflow State & Disconnected / linear & Unified / iterative \\ \bottomrule
\end{tabular}}
\end{table}

This paper introduces \textbf{Quality Audio Prototyping (QuAP)}, a working prototype application developed in JUCE \footnote{\url{https://juce.com}} that unifies sound retrieval and procedural generation within a single interface. QuAP addresses the friction of the exploratory design phase through three integrated components:

\begin{itemize}
    \item\textbf{Hybrid Retrieval:} Combines text-based metadata search with content-based similarity retrieval, using a MobileNet architecture deployed via ONNX Runtime to extract deep acoustic embeddings, indexed and queried using FAISS\cite{johnson2019billion} for near-instantaneous results.
    
    \item \textbf{Procedural Synthesis:} Embeds six optimised  synthesis models: spanning physical, modal, and subtractive approaches, enabling the generation of new sounds from scratch.
    
    \item \textbf{Intelligent Guidance:} An embedded rule-based assistant provides perceptually informed parameter recommendations for each procedural model, indicating the ranges within which synthesis parameters were perceived as most realistic by participants in a subjective evaluation. This lowers the barrier to procedural interaction and supports rapid creative exploration without removing designer agency.
\end{itemize}

By unifying similarity-based retrieval, procedural generation, and hybrid layering within a single environment, QuAP addresses the procedural distance between a narrative concept and its sonic realisation. Preliminary validation confirms the viability of this approach: the optimisation of procedural models demonstrated statistically significant quality improvements in five of the six embedded categories ($p < 0.05$); the ablation study established MobileNetV3 as the preferred encoder for sound effect retrieval, outperforming a ResNet18-IBN baseline on mean average precision (mAP: 0.449 vs. 0.412) on the FSD50K dataset; and a user evaluation with 16 practitioners confirmed the tool's workflow utility, with all participants agreeing that the parameter assistant preserved creative agency within an increasingly tool-fragmented workflow.

\section{Related Work}
\label{ssec:previouswork}
The evolution of sound design tools has developed along two primary axes: the ease of retrieval from large-scale audio databases, and the flexibility of real-time sound synthesis. In professional practice, these two axes remain largely disconnected. Practitioners typically navigate between separate retrieval platforms, synthesis  tools, generative solutions, and organisational tools to construct a single workflow. A significant limitation of existing generative and retrieval tools is their predominant orientation toward music production. The majority of large-scale audio models are trained on datasets with strong musical bias thus prioritising tonal, rhythmic, and melodic structure~\cite{MusicLM, liu2023audioldm}. This music-centric bias poorly generalises to the nonlinear, texture-based, and physically motivated characteristics of sound effects and environmental audio~\cite{Camara}. Therefore this bias represents a critical gap for sound design practitioners, whose requirements centre on parametric control, timbral specificity, and narrative fit rather than musical coherence. Commercial tools such as Splice\footnote{\url{https://splice.com}}, 
Krotos\footnote{\url{https://krotos.com}}, ElevenLabs\footnote{\url{https://elevenlabs.com}}, and 
iZotope\footnote{\url{https://www.izotope.com}} address individual aspects of this pipeline but do not offer an integrated solution. QuAP is positioned as an attempt to unify these axes within a single environment, reducing the context-switching that interrupts the exploratory phase of sound design work. 
 
\subsection{Content-Based Audio Retrieval }

Content-based audio retrieval refers to the task of retrieving audio recordings based on their acoustic characteristics of a query audio clip. The broader literature encompasses several related problem formulations. High-specific audio fingerprinting systems aim to identify near-identical matches within large databases \cite{wang2003industrial, chang2021neural}. Closely related tasks include music version identification, which detects different renditions of the same musical work \cite{serra2009cross, yesiler2020accurate}, and query-by-vocal-imitation, where users retrieve sounds by producing vocal imitations of the target audio \cite{blancas2014sound, zhang2016imisound}. 

Recent work has increasingly approached these problems through learned audio embeddings that map sounds into a metric space where semantically similar audio segments are located nearby. Convolutional neural networks trained for large-scale audio tagging have proven particularly effective as general-purpose audio encoders \cite{kong2020panns}. In particular, lightweight architectures such as MobileNet \cite{howard2017mobilenets} provide a favorable balance between representational capacity and computational efficiency, making them well suited for real-time and on-device audio retrieval applications. Given its utility in related tasks \cite{greif2024improving, Garcia2025qbvi}, we use MobileNet as the encoder architecture for our audio embedding system. 

\begin{table*}[h]
    \centering
    \resizebox{\textwidth}{!}{
    \begin{tabular}{llll}
        \toprule
        \textbf{Model} & \textbf{Synthesis Type} & 
        \textbf{Parameters} & \textbf{Optimisation} \\
        \midrule
        Fire & Additive, modal, subtractive & Lapping, hissing, crackling, intensity & Reverb, compression and EQ in low frequencies \\
        Explosion & Additive, modal, physically informed & Rumble, air, dust (decay, amount) & Reverb, distortion and compression \\
        Jet & Additive, physically informed & Turbine, burn, speed & Reverb, distortion and EQ in high frequencies \\
        Rocket & Subtractive, modal & Duration, chamber resonance & Reverb and compression \\
        Helicopter & Physically informed & Period, frequency, distance & Reverb and distortion \\
        Gun & Additive, modal, physically informed & Shell frequency, shell decay & Reverb and compression \\
        \bottomrule
    \end{tabular}}
    \caption{Procedural audio models embedded in QuAP, with 
    synthesis type, exposed parameters, and applied 
    optimisations.}
    \label{tab:ProcAudio}
\end{table*}

\subsection{Procedural Audio and Intelligent Interfaces}

Parallel to retrieval, procedural audio offers a dynamic alternative to static samples. By employing mathematical models to simulate physical or electronic sound-generating processes, designers can achieve a degree of parametric variability and real-time control that sample-based approaches cannot provide~\cite{farnell2010designing}. In this sense, procedural audio functions as a form of rule-based synthesis (analogous to digital Foley) in which timbral and temporal characteristics are governed by adjustable parameters rather than fixed recordings.

However, the high dimensionality of synthesis parameter spaces presents a significant barrier to adoption. Recent work in intelligent sound engineering has explored mapping complex synthesis parameters to lower-dimensional, perceptually meaningful representations, making procedural tools more accessible to practitioners without deep technical expertise~\cite{Dimitris}. QuAP main design is presented in Table \ref{tab:workflow_comparison} it builds on this direction by embedding a parameter assistant that constrains interaction to perceptually validated ranges, derived from the optimisation process described in Section~\ref{ssec:methodology}.

In professional practice, output quality is not always the primary objective, a sound that serves the narrative and fits the production context is often sufficient, particularly under deadline pressure. Current procedural models are therefore well-suited as a foundation for layering and post-processing, rather than as a replacement for recorded material~\cite{Kamath}. Additionally as AI-driven tools become increasingly prevalent in creative workflows, practitioners require greater transparency and control over their assets, particularly regarding the provenance of training data and the legibility of generative outputs~\cite{Herington}.

\section{Methodology}
\label{ssec:methodology}

The development of QuAP required the optimisation of two independent but complementary components: the procedural audio models that underpin the synthesis engine, and the audio embedding model that drives similarity-based retrieval. For the procedural models, optimisation was guided by a feature-driven bottleneck framework \cite{koh2020conceptbottleneckmodels}, which identifies the acoustic features most salient for perceptual classification, with results validated through subjective evaluation. For the retrieval component, encoder architecture selection was informed by performance on content-based audio retrieval benchmarks, detailed in the ablation study presented in Section~\ref{AblationSection}. Together, these methodological decisions define the technical foundation of the integrated system described in Section~\ref{ssec:design}.

\section{System Design}
\label{ssec:design}
\begin{figure*}
  \centering
  \includegraphics[width=0.8\linewidth]{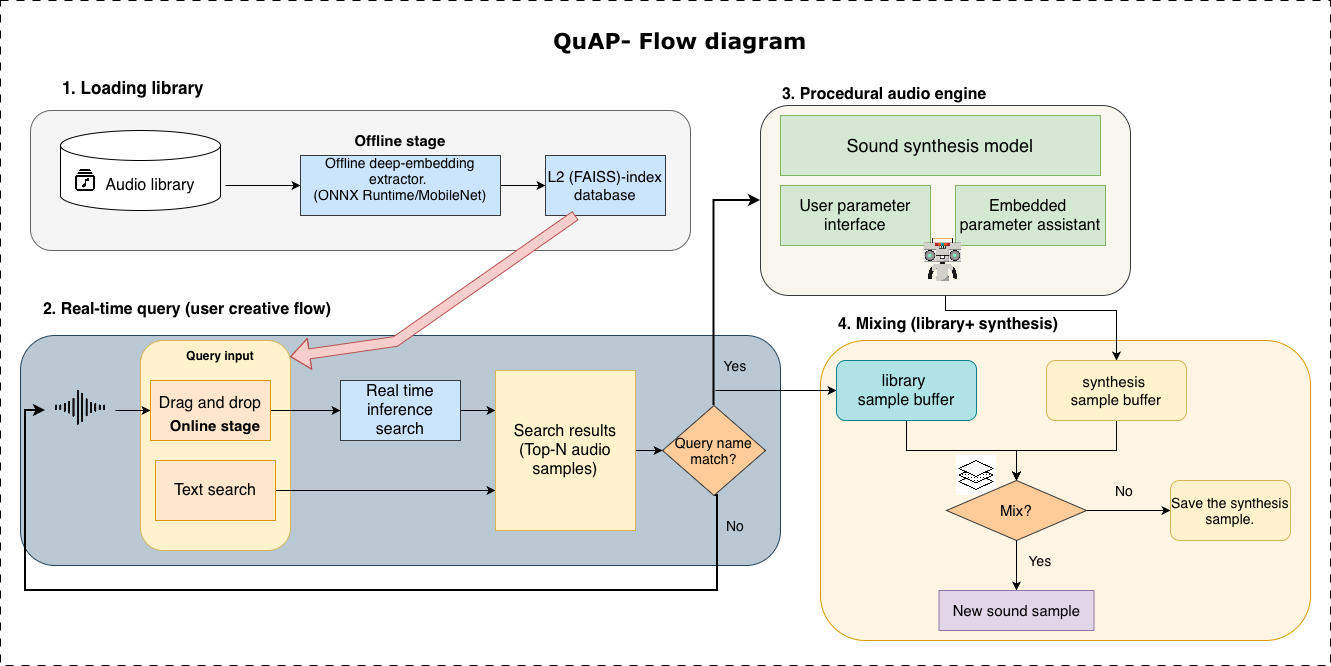}
 \caption{QuAP system architecture. The loading library stage (top left) processes the user's audio library through an offline deep-embedding extractor and FAISS indexing backend, executed asynchronously in a background thread. The real-time query stage (bottom left) accepts drag-and-drop or text input, performing similarity search against the indexed database. Where procedural models are available, the system routes to the procedural audio engine (top right) for parameter-based sound generation. The resulting synthesis output may then be layered with the retrieved library sample in the mixing stage (bottom right), allowing the designer to blend static and dynamic material within a single interface.}
 \label{QuAPdiagram}
\end{figure*}

The system integrates four primary components: (1) an offline embedding and indexing pipeline, (2) a real-time query inference module, (3) an interface for interaction with procedural models, and (4) a hybrid layering stage enabling the user to combine retrieved library samples with procedurally generated audio. An overview of the system architecture is presented in Figure~\ref{QuAPdiagram}.

The system operates across two stages. In the \textbf{offline stage}, audio samples from the user's library are processed by a deep embedding model to produce fixed-dimensional feature representations, which are stored in a FAISS-indexed database. To maintain plugin responsiveness, this extraction and indexing process is executed asynchronously in a background thread, with progress notifications surfaced in the interface. In the \textbf{online stage}, a query audio clip is processed through the same embedding model, and the resulting embedding is compared against the indexed database to retrieve acoustically similar samples. More details are discussed in the next subsections. 

\subsection{Procedural Audio Models}

QuAP embeds six procedural audio models, available through the Nemisindo synthesis engine\footnote{\url{https://nemisindo.com}}, and inspired by the design principles of Farnell~\cite{farnell2010designing}. The models, their synthesis types, exposed parameters, and applied optimisations are summarised in Table~\ref{tab:ProcAudio}, classified following~\cite{Dimitris}.

\subsection{Procedural Model Optimisation}
\label{ProceduralModelOptimisation}
\begin{figure*}[h]
    \centering
    \includegraphics[width=0.8\linewidth]
    {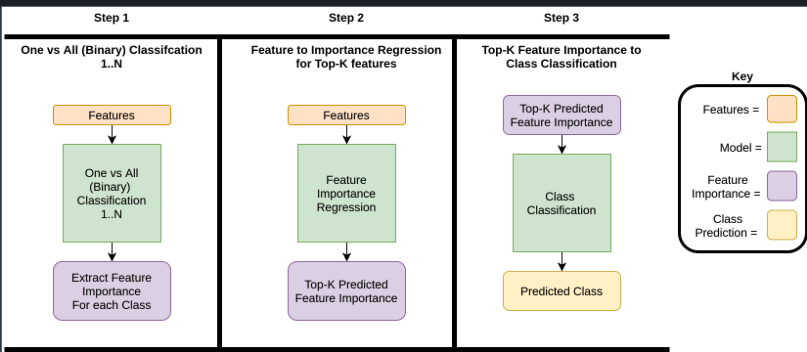}
    \caption{Feature-driven bottleneck framework used to optimise procedural audio model parameters. Steps 1--3 correspond to one-versus-all classification, top-$K$ feature importance regression, and top-$K$ feature  classification respectively.}
    \label{fig:bottleneckProcess}
\end{figure*}

A central design challenge in embedding procedural models within a creative tool is determining which synthesis parameters are perceptually meaningful. That is, which low-level acoustic features, when modified, are most likely to influence how a user perceives the output. To address this, we applied a feature-driven bottleneck framework~\cite{koh2020conceptbottleneckmodels} to the six sound categories embedded in QuAP. Using low-level audio features extracted via Essentia~\cite{essentia} from the 6KSFX dataset~\cite{6ksfx}, the framework identifies the top-$K$ acoustic features that most reliably discriminate between recorded and synthetic audio for each category, as illustrated in Figure~\ref{fig:bottleneckProcess}. Based on these feature importance results, targeted post-production effects were applied to each model, and the exposed parameter ranges recommended by the embedded assistant were derived from these findings. The optimisations applied to each category are listed in Table~\ref{tab:ProcAudio}.

A MUSHRA subjective evaluation~\cite{MUSHRA1} with 20 participants was conducted to validate the optimisations. The study was conducted in accordance with institutional ethical guidelines. Participants provided informed consent prior to participation, and all data were anonymised before analysis. Statistically significant quality improvements were observed in five of the six categories ($p < 0.05$), as summarised in Table~\ref{tab:mushra}. The Rocket model did not reach statistical significance ($p = 0.08$), indicating that post-production effects alone are insufficient for this category and that synthesis-level modifications may be required. The Jet category was excluded from the final subjective evaluation, as participants consistently noted that the optimised outputs remained perceptually too synthetic relative to recorded reference material. Full optimisation details are documented at the project website\footnote{\url{https://saop-project.netlify.app}} and in a companion study currently under review~\cite{Garcia2025bottleneck}.

\begin{table}[h]
\centering
\caption{MUSHRA subjective evaluation results for the six procedural models embedded in QuAP. Default refers to the unoptimised synthetic sample. Best Opt. reports the highest mean score across optimisation variants.}
\label{tab:mushra}
\resizebox{\linewidth}{!}{
\begin{tabular}{lcccc}
\toprule
\textbf{Model} & \textbf{Default} & \textbf{Best Opt.} & 
\textbf{F} & \textbf{p} \\
\midrule
Fire       & 28.85 & 40.45 & 15.23 & $<0.001$* \\
Explosion  & 56.40 & 52.55 & 11.39 & $<0.001$* \\
Helicopter & 41.10 & 51.20 & 28.74 & $<0.001$* \\
Rocket     & 37.85 & 49.20 & 1.75  & $0.08$ \\
Gun        & 35.95 & 45.60 & 3.54  & $<0.001$* \\
Jet        & \multicolumn{4}{c}{Excluded — see limitations} \\
\bottomrule
\multicolumn{5}{l}{* $p < 0.05$}
\end{tabular}}
\end{table}

\subsection{Audio Embedding Model}
\label{ssec:encoder}

Audio similarity is modeled using a convolutional embedding network based on the MobileNet architecture \cite{howard2017mobilenets}. The backbone is initialized using weights pre-trained for large-scale audio tagging on the AudioSet \cite{gemmeke2017audioset} dataset using the EfficientAT framework \cite{schmid2023efficient}. These models are trained to perform multi-label sound event classification and provide strong general-purpose audio representations for downstream tasks. 

The network is further fine-tuned using supervised contrastive learning \cite{khosla2020supervised}, which encourages embeddings from the same semantic class to cluster together while separating embeddings from different classes in the latent space. This is done using the FSD50K dataset \cite{fonseca2021fsd50k} which provides hierarchical class labels following the AudioSet ontology.

\subsection{Offline Embedding Extraction and Indexing}

When a user loads a sound library into QuAP, each audio file is processed through the embedding network to generate a fixed-dimensional vector representation. These embeddings are computed offline and stored in a FAISS \cite{johnson2019billion} vector database that supports fast nearest-neighbor search. This indexing step significantly reduces the computational cost of subsequent similarity queries. Because sound libraries may contain thousands of audio files, the embedding extraction and indexing procedures are executed asynchronously in a background thread. The user interface provides progress feedback during this process.

\subsection{Real-Time Query and Retrieval}

For similarity search, users can drag and drop an audio sample into the plugin interface. The query sample is processed through the embedding network to generate a query embedding vector. The system then performs a nearest-neighbor search against the indexed embedding database to retrieve the most acoustically similar samples. The retrieved results are ranked according to embedding similarity and displayed within the plugin interface. 

To ensure compatibility with different hardware configurations, the system performs a lightweight calibration step at initialization. This step estimates the computational capacity of the host machine and adjusts inference parameters to maintain real-time performance.

\subsection{Plugin Interface and Procedural Interaction}
The interface dynamically adapts when procedural audio models are available for certain sound categories. In these cases, the plugin exposes additional parameter controls that allow users to manipulate the procedural sound generation model directly. The embedded assistant also recommends optimal parameter ranges whenever available, based on known perceptual knowledge \cite{Garcia2025bottleneck} . This design allows QuAP to function both as a sound retrieval system and as an exploratory sound design assistant.

\begin{figure}
    \centering
    \includegraphics[width=0.8\linewidth]{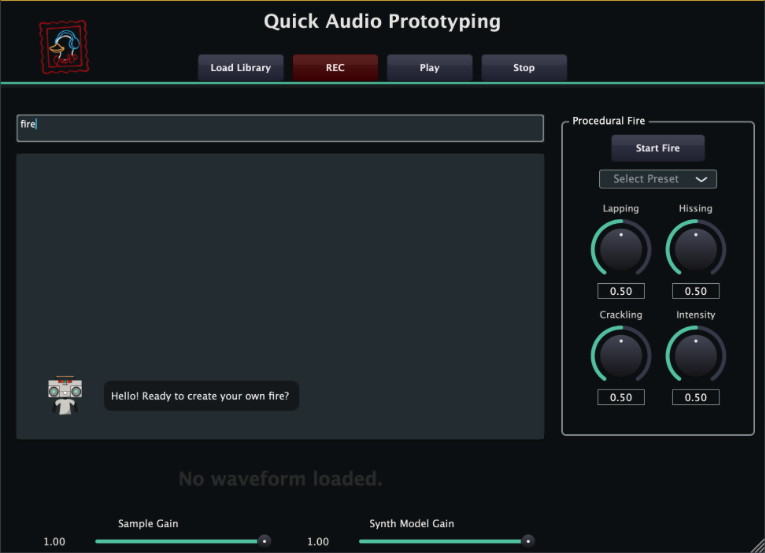}
    \caption{QuAP graphical user interface. The left panel displays the audio library search results, supporting both text-based and drag-and-drop similarity queries. The right panel exposes the procedural audio model controls, with the embedded assistant providing perceptually informed parameter recommendations.}
    \label{fig:GUI}
\end{figure}

An example of the GUI is illustrated \ref{fig:GUI}, where the procedural audio panel, the embedded assistant and the library panel can be seen.  

\subsection{Embedded Parameter Assistant}

The embedded parameter assistant is implemented as a rule-based guidance system, distinct from large language models or generative AI. For each of the six procedural models embedded in QuAP, the assistant serves two complementary functions. First, it displays pre-defined parameter ranges derived directly from the feature-driven bottleneck optimisation described in Section~\ref{ssec:methodology}, specifically, the configurations under which participants in the MUSHRA subjective evaluation reported the highest perceptual similarity to recorded reference material. Second, it provides a plain-language description of what each synthesis parameter controls, allowing practitioners to understand the perceptual effect of a slider without requiring prior knowledge of the underlying synthesis model. This addresses a known barrier to procedural audio adoption~\cite{Menexopoulos}, where the high dimensionality and technical naming of synthesis parameters often discourages non-expert users. The assistant does not generate or adapt recommendations dynamically; rather, it surfaces empirically validated ranges and parameter descriptions as guidance overlaid on the synthesis controls, allowing the designer to make informed adjustments while retaining full creative 
control over the output.

\section{User Evaluation}
\label{Participants}
\begin{table}[t]
\centering
\caption{Ablation study comparing encoder architectures for audio retrieval. }
\label{tab:encoder_ablation}
\begin{tabular}{lcc}
\hline
\textbf{Encoder} & \textbf{mAP} $\uparrow$ & \textbf{NDCG} $\uparrow$ \\
\hline
ResNet18-IBN & 0.412 & 0.625 \\
MobileNetV3 (ours) & \textbf{0.449} & \textbf{0.656} \\
\hline
\end{tabular}
\end{table}

Ethical approval for this study was obtained in accordance with institutional guidelines. All participants provided informed consent prior to taking part, were informed of their right to withdraw at any time, and all responses were anonymised before analysis.

QuAP was evaluated by 16 participants following up their interest in the survey: 8 sound designers, 4 audio researchers, and 4 music producers. Evaluation sessions were conducted both in-person and remotely. In-person sessions employed a cognitive walkthrough methodology~\cite{blandford2016qualitative}, in which the evaluator observed how participants interacted with the interface during a guided task. The primary task required participants to locate a fire sound sample within their library using the similarity search, and subsequently generate a new variant using the procedural audio model. For remote sessions, participants were provided with the standalone plugin, a written instruction page, and a video tutorial \footnote{\url{https://quap.netlify.app}}. 

Participants were asked to assess workflow utility, deployment context, and suggested improvements. 75\% of the participants found QuAP useful and potentially helpful in their workflows. The plugin and supporting materials are available in the supplementary documents accompanying this paper.

Transcripts from all sessions were analysed using thematic analysis~\cite{braun2006using}, yielding five themes and thre relevant codes are presented in Table~\ref{tab:themes}. \textit{Workflow Integration} captures how participants positioned QuAP within their existing pipeline, particularly regarding library navigation and the identification of unmet tooling needs. Participants highlighted the hybrid layering feature as a notable strength, enabling sound samples to 
be augmented directly with procedural model parameters within a single environment. \textit{Interface} reflects responses to the usability and responsiveness of the plugin, with participants suggesting improvements including a visible list of available procedural models and greater prominence of the embedded assistant within 
the interface. \textit{Creative Agency} encompasses the extent to which participants felt the tool supported rather than replaced their creative decision-making, particularly the role of the parameter assistant in maintaining human control over synthesis outputs. 
\textit{Quality \& Assessment} addresses how participants evaluated the procedural models: quality concerns were minimal, with participants instead emphasising the 
utility of layering synthetic outputs with recorded library material as a practical creative strategy. 
\textit{Future Potential} captures expressed interest in extending the tool's capabilities, including broader model coverage and deeper integration within production 
workflows.
\begin{table*}[h]
\centering
\caption{Thematic framework derived from user evaluation responses.}
\label{tab:themes}
\begin{tabular}{lll}
\toprule
\textbf{Theme} & \textbf{Code} & \textbf{Description} \\
\midrule
\multirow{3}{*}{Workflow Integration} & Management & Library organisation and asset handling \\
& Similarity & Acoustic search and retrieval relevance \\
    & Areas of opportunity & Identified gaps in current tooling \\
\midrule
\multirow{2}{*}{Interface} 
    & GUI & Interface usability and visual feedback \\
    & Computational Process & System performance and responsiveness \\
\midrule
\multirow{4}{*}{Creative Agency} 
    & Human-in-the-loop & Designer control over generative output \\
    & Augmentation & AI as creative support, not replacement \\
    & Creation & Subjective aesthetic judgement \\
    & Narrative & Fit of sound within storytelling context \\
\midrule
\multirow{3}{*}{Quality \& Assessment} 
    & Quality & Perceived realism of synthesis models \\
    & Assessment & Evaluation of output quality \\
    & Limitation & Current scope and implementation gaps \\
\midrule
\multirow{4}{*}{Future Potential} 
    & Potential & Perceived future utility of the tool \\
    & Possibilities & Suggested extensions and new use cases \\
    & Waveform & Visual audio feedback and representation \\
    & Storyboard & Integration within broader production pipeline \\
\bottomrule
\end{tabular}
\end{table*}

\section{Ablation Study}
\label{AblationSection}
The choice of audio embedding architecture is guided by two design considerations: the model should be lightweight enough for efficient deployment, while maintaining strong performance on related content-based audio retrieval tasks. Our primary encoder is based on MobileNetV3, motivated by the findings of Greif et al.~\cite{greif2024improving}, who demonstrate that fine-tuning MobileNetV3 yields state-of-the-art performance in query-by-vocal-imitation retrieval.

As a baseline, we employ a ResNet18-IBN \cite{du2021bytecover} encoder. This architecture represents a lightweight variant of convolutional backbones previously used in tasks such as cover song identification \cite{du2021bytecover} and music sample identification \cite{bhattacharjee2025refining}. To ensure a fair comparison, the baseline model is trained using the same data preprocessing and training procedure as the MobileNet-based encoder described in Section~\ref{ssec:encoder}.

Both models are evaluated on a held-out test split of the FSD50K dataset. Retrieval performance is measured using mean average precision (mAP) and normalized discounted cumulative gain (NDCG). Table \ref{tab:encoder_ablation} summarizes the comparative performance of the two architectures.

\section{Discussion}

The results across the three evaluation components — procedural, model optimisation, encoder ablation, and user evaluation — collectively support the core premise of QuAP: that unifying retrieval and synthesis within a single, interpretable environment is both technically viable and practically valuable for sound design workflows.

\subsection{Procedural Model Quality and Perceptual Utility}

The MUSHRA evaluation results presented in 
Section~\ref{ProceduralModelOptimisation} reveal an important distinction between perceptual realism and workflow utility. While several models, most notably Fire (best opt: 40.45) remain within the 'slight resemblance' range of the MUSHRA scale, this does not preclude their practical value in a professional context. As confirmed by the user evaluation  in section \ref{Participants}, practitioners do not always require acoustically perfect outputs; a sound that serves the narrative and provides a viable foundation for layering is often sufficient, particularly during the exploratory phase of a project.

The Rocket model's failure to reach statistical significance ($p = 0.08$) indicates that post-production effects alone are insufficient for certain synthesis categories, and that meaningful improvement would require modifications at the synthesis algorithmic level. This is an acknowledged limitation of the current implementation, and highlights a broader challenge in procedural audio optimisation: the ceiling of post-production enhancement is bounded by the quality of the underlying synthesis model.

The Explosion category presents a notable result: the best optimisation (52.55) performed below the default synthetic sample (56.40), suggesting that the applied post-production effects introduced perceptual artefacts rather than improvements for this category. This underscores the importance of category-specific optimisation strategies rather than uniform application of effects.

The feature-driven bottleneck framework provided a further contribution beyond model improvement: interpretability. By identifying which acoustic features most significantly differentiate real from synthetic audio, the framework 
generates actionable insight for both developers and practitioners. These insights directly informed the parameter assistant embedded in QuAP's interface, which 
constrains interaction to perceptually validated ranges , translating machine-derived feature importance into human-readable guidance that supports critical listening 
rather than delegating decisions to automation.

\subsection{Encoder Selection and Retrieval Performance}

The ablation study confirms that MobileNetV3 outperforms the ResNet18-IBN baseline on both mAP (0.449 vs 0.412) and NDCG (0.656 vs 0.625) on the FSD50K held-out test split. While the margin is modest, the choice of MobileNetV3 is additionally motivated by its computational efficiency — a critical consideration for real-time deployment within a  Digital audio work station environment where latency and  random access memory (RAM) constraints are non-trivial. The use of FSD50K, a sound event dataset, as the fine-tuning corpus is a deliberate departure from music-centric training data, ensuring that the embedding space is optimised for the nonlinear, texture-based characteristics of sound effects rather than tonal or rhythmic structures.

\subsection{Human-in-the-Loop Design}

The user evaluation finding that all participants found the embedded assistant reduced the barrier to procedural interaction, with one noting it \textit{"gives the sensation of keeping human-in-the-loop"}, reflects a broader design principle that guided QuAP's development. The goal was not to automate sound design, but to make the technical complexity of synthesis accessible without 
removing the designer's creative agency. By exposing perceptually meaningful parameters and providing guidance grounded in empirical feature importance, QuAP encourages practitioners to engage with and develop their critical listening, using the tool as a creative instrument rather than a generative shortcut.

The four participants who identified limitations citing scope constraints or unaddressed bottlenecks which highlighted the boundaries of the current pilot implementation. 
QuAP currently supports six sound categories; expanding the procedural model library and improving the synthesis quality of categories like Rocket and Jet represents the primary direction for future development.

\section{Conclusion}

This paper introduced Quality Audio Prototyping (QuAP), a working prototype application that unifies content-based audio retrieval and procedural sound generation within a single environment. By combining a MobileNetV3-based similarity search engine with six optimised procedural audio models and an embedded parameter assistant, QuAP addresses the workflow fragmentation that characterises current sound design practice.

Three complementary evaluations validate the system. The MUSHRA subjective evaluation confirmed statistically significant quality improvements in five of the six embedded sound categories following feature-driven optimisation, demonstrating that procedural models ( while not acoustically perfect) are perceptually viable as a foundation for layering and creative exploration. The ablation study established MobileNetV3 as the preferred encoder for sound effect retrieval, outperforming the ResNet18-IBN baseline on both mAP and NDCG on the FSD50K dataset. The user evaluation confirmed practitioner interest in the integrated approach, with 10 of 16 participants identifying QuAP as a viable workflow tool, and all participants agreeing that the parameter assistant lowered the barrier to procedural interaction while preserving creative agency.

The central contribution of QuAP is not the replacement of any single existing tool, but the unification of retrieval and synthesis within an interpretable, human-in-the-loop environment. To create a tool that gives designers knowledge of what each parameter means and the creative space to explore it. Future work will focus on expanding the procedural model library, improving synthesis quality for categories that did not reach statistical significance, and conducting a larger-scale user study to assess long-term workflow integration.

\bibliographystyle{IEEEtranDAFx}
\bibliography{DAFx26_tmpl} 

\end{document}